\documentclass[aps,pra,superscriptaddress,reprint, floatfix]{revtex4-1}
\usepackage{amstext,amssymb}
\usepackage{amsmath}
\usepackage{amsfonts}
\usepackage{xcolor}
\usepackage{graphicx}
\usepackage{bbold}
\usepackage[T1]{fontenc}
\def\Tr{\mbox{Tr}}
\newcommand{\ini}{\varrho} 
\begin{document}
\title{To Measure, or Not to Measure, That is the Question}
\author{Juzar Thingna}
\email[]{jythingna@ibs.re.kr}
\affiliation{Center for Theoretical Physics of Complex Systems, Institute for Basic Science (IBS), Daejeon 34126, Republic of Korea}
\author{Peter Talkner}
\email[]{peter.talkner@physik.uni-augsburg.de}
\affiliation{Institut f\"{u}r Physik, Universit\"{a}t Augsburg, Universit\"{a}tsstrasse 1, D-86135 Augsburg, Germany} 
\affiliation{Center for Theoretical Physics of Complex Systems, Institute for Basic Science (IBS), Daejeon 34126, Republic of Korea}
\date{\today}
 
\begin{abstract}
A method is proposed that allows one to infer the sum of the values of an observable taken during contacts with a pointer state. Hereby the state of the pointer is updated while contacted with the system and remains unchanged between contacts while the system evolves in time. After a prescribed number of such contacts the position of the pointer is determined  by means of a projective measurement. The outcome is specified in terms of a probability distribution function for unitary and Markovian dissipative dynamics and compared with the results of the same number of generalized Gaussian measurements of the considered observable. As a particular example a qubit is considered with an observable contacting to the pointer that does not commute with the system Hamiltonian.   
\end{abstract}   
\maketitle
\section{Introduction}\label{I}
Measurements play an important role in science in general and in quantum mechanics in particular. While in classical systems measurements can in principle be performed with unlimited precision and without any influence on the measured object, for quantum systems often there are principle limits of the achievable precision and unavoidable, sometimes drastic back-actions on the state of the measured object. The frequent repetition of the same measurement may either lead to the total freezing of the system's dynamics, known as Zeno effect \cite{Sudarshan77} or to a steady heating of the system \cite{Yi, Magazzu}, effects that are alien to classical systems. Because the only way of gaining information about the state of a quantum system is by measuring, understanding the measurement process and its impact on the considered system is vital. From the point of view of a theoretician, projective measurements, wherein the system state collapses to the measured state \cite{Neumann18},  are most convenient. This idea of projective measurements leads to simplistic theoretical approaches but lacks information about the measuring device and its properties as well as about possible deviations from the ideal picture.

\begin{figure}[t]
\includegraphics[width=\columnwidth]{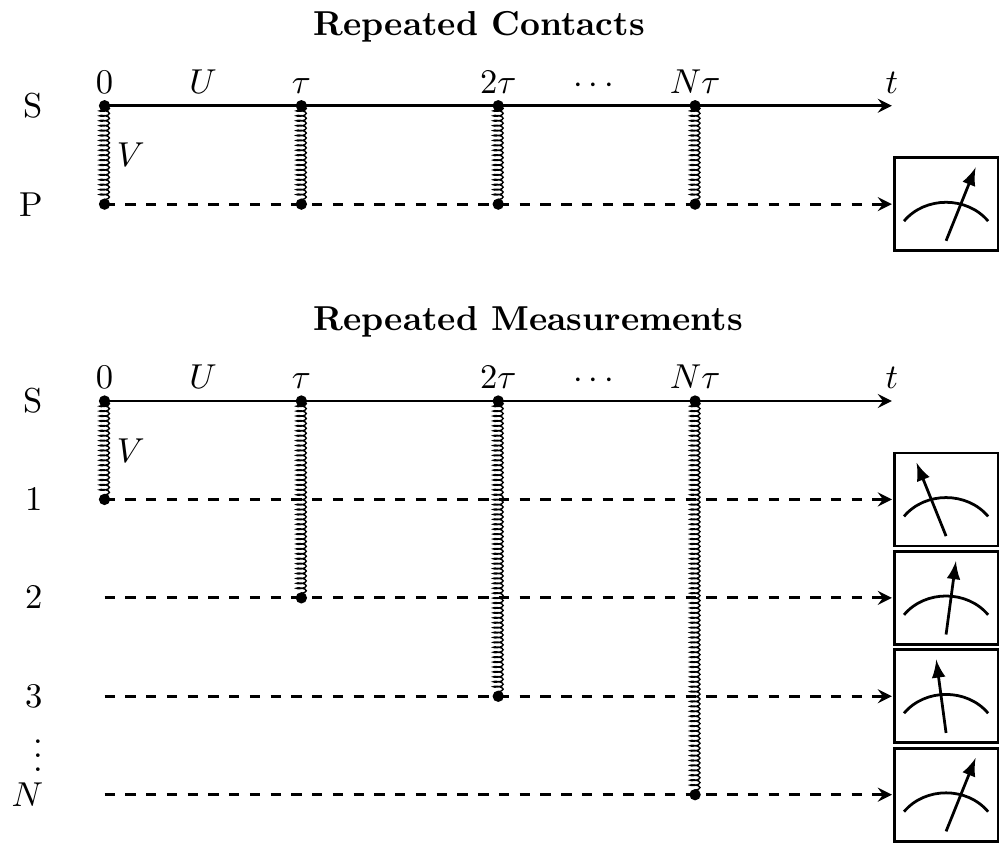}
\centering
\caption{\label{fig:0}Schematic illustration of the repeated contact and repeated measurement schemes. The solid lines represent the system evolution whereas the dashed lines correspond to the pointer evolution. The system is periodically connected with the pointer at intervals of duration $\tau$. The repeated contact scheme is  less invasive as compared to the repeated measurements. The latter provides information about the full ``trajectory'' of measured values while the former only allows one to infer about their sum.}
\end{figure}

Alternatively, as shown in this work, one could adapt von Neumann's projective measurement approach to generalized measurements in which the measuring device is a quantum object that comes in \emph{contact} with the system. During contact both system and the device affect each other and hence by projectively measuring the device after contact one can infer information about the system \cite{Neumann18,Bohm,Wiseman,Talkner16}. Here we explore the possibility of repeated contacts of the measuring device to record information from the system which would be read out at the very end. We compare the proposed $N$ times repeated contacts approach to the $N$ times repeated measurements case where after each contact the measuring device is read out (see illustration in Fig.~\ref{fig:0}).

\section{A single generalized measurement}\label{sgm}
Following von Neumann's approach \cite{Neumann18,Bohm}, we consider a quantum measuring device called ``pointer'' that comes in contact with the system for a short time $\tau_p$ with strength $g$. The contact time is extremely short as compared to the timescale of system dynamics, such that, the system does not evolve during the contact. Thus, whenever the pointer connects to the system, the density matrix $\rho_{tot}$ of the combined system pointer immediately before the contact is modified to the post-contact state $\tilde{\rho}_{tot}$ according to
\begin{eqnarray}
\label{eq:1}
\tilde{\rho}_{tot} &= V \rho_{tot} V^{\dagger}\:,
\end{eqnarray}
where the unitary time evolution operator $V$ being determined by the action of the system-pointer interaction Hamiltonian $H_{SP} = g M P$ is given by
\begin{equation}
  V = e^{-i \kappa M P/\hbar}\:.
  \label{eq:3}
\end{equation}
Here, the system operator $M$ represents the observable to be measured and $P$ is the momentum operator which is conjugate to the pointer position operator $Q$; finally, $\kappa = g \tau_p$ is an effective measure of the interaction strength. As an operator being exponential in the pointer momentum operator $P$, $V$ shifts the pointer position by an amount depending on the state of the system. If immediately after the system pointer contact a projective measurement of the pointer state with respect to the position $x$ is performed, the non-normalized density matrix of the system follows by the action of the operation $\phi^1_x$~\cite{op,Kraus,Hayashi} which is given by
\begin{equation}
  \begin{split}
    \phi^1_x( \ini) & = (x |\tilde{\rho}_{tot}|x) \\
    &= \sum_{m,m'} \mathcal{P}_m \ini \mathcal{P}_{m'} \sigma(x-\kappa \mu_m,x-\kappa \mu_{m'})\:,
  \end{split}
  \label{fi1}
\end{equation}
where we assumed that the density matrix of the total system $\rho_{tot}$ factorizes in a direct product of density matrix of the system $S$, $\ini$, and that of the pointer, $\sigma$. Here, $|x)$ is the eigenstate of the pointer position operator belonging to the eigenvalue $x$, and $e^{- i a P/\hbar}|x ) =|x+ a )$. Further, $\mathcal{P}_m$ is the projection operator of the system observable $M$ onto the subspace belonging to the eigenvalue $\mu_m$ and the position matrix elements of the pointer density matrix are denoted by
\begin{equation}
  \sigma (x,y) = (x|\sigma |y )\:.
  \label{sxy}
\end{equation}
The probability density function (pdf) $P_1(x)$ with which the pointer position $x$ is observed is determined by trace of the operation acting on the system density matrix yielding
\begin{equation}
  \begin{split}
    P_1(x) &= \Tr\, \phi^1_x(\ini)\\
    &= \sum_m p_m \sigma(x-\kappa \mu_m,x-\kappa \mu_m)\:, 
\label{p1}
  \end{split} 
\end{equation} 
where 
\begin{equation}
p_m =\Tr\, \mathcal{P}_m \ini 
\label{pm}
\end{equation}
represents the probability to find a system with density matrix $\ini$ in the subspace belonging to the eigenvalue $\mu_m$. For a pure Gaussian pointer state with vanishing mean value of the position and the momentum and variance $\langle Q^2 \rangle$, the density matrix takes the form
\begin{equation}
\sigma(x,y) = \frac{1}{\sqrt{2 \pi \langle Q^2 \rangle }}e^{-(x^2+y^2)/(4 \langle Q^2 \rangle)}\:.
\label{sig}
\end{equation}
Then the  pdf $P_1(x)$ becomes a mixture of Gaussians $g_{\langle Q^2 \rangle}(x- \kappa \mu_n)$ with weights $p_m$ where $g_v(x)=(2 \pi v)^{-1/2} \exp \{-x^2/(2v)\}$ denotes a Gaussian pdf with vanishing mean value and variance $v$. By rescaling the pointer variable according to $x=\kappa\, \mathsf{x}$, the maxima of the accordingly transformed pdf $P_1(\mathsf{x})$ are shifted towards the eigenvalues $\mu_m$. The resulting scaled pdf hence becomes
\begin{equation}
P_1(\mathsf{x}) = \sum_m p_m\, g_{\sigma^2_\mathsf{x}}(\mathsf{x} -\mu_m)\:.
\label{sp1}
\end{equation}
In the mixture~\eqref{sp1} those maxima survive whose weights are sufficiently large and for which the rescaled variance $\sigma^2_\mathsf{x} = \langle Q^2 \rangle /\kappa^2$ is sufficiently smaller than the squared smallest distance between the eigenvalues.

\section{Multiple contacts}\label{mc}
Rather than to consider the statistics of the outcomes of a single measurement of the considered observable $M$ we are asking for the statistics of the sum of $N$ values of the observable $M$ that are assumed at subsequent times. The registration of the observable can be realized in different ways which in general lead to different results. A straightforward procedure is to repeat the above described generalized measurement $N$ times always after the time $\tau$ has elapsed, as sketched in the lower panel of Fig.~\ref{fig:0}. We shall come back to this approach later. First we follow the strategy illustrated in the upper panel of Fig.~\ref{fig:0}. This approach consists in $N$ repetitions of contacts acting via the unitary operator $V$ on the composed system each followed by a unitary time-evolution $U$ of the system alone during a time $\tau$ while the pointer remains unaffected. The time evolution of the system from a contact to the next one is governed by the Hamiltonian $H_S$ and hence given by 
\begin{equation}
U = e^{-i H_S \tau/\hbar}.
\label{U}
\end{equation}  
The action of $N$ combined contacts and time evolutions on the total, initially factorizing, density matrix is then given by
\begin{equation}
\rho_{tot}(N \tau) = (U V)^N \ini \otimes \sigma (V^\dagger U^\dagger)^N\:.
\label{UVN}
\end{equation}        
In analogy to the case of a single measurement, after the completion of the $N$ contact protocol one may read out the pointer state by a projective measurement. The non-normalized reduced density matrix conditioned on the measured result $\mathsf{x}$ is determined by the operation $\phi^N_\mathsf{x}$ which acts as
\begin{equation}
\begin{split}
\phi^N_\mathsf{x}(\ini) &= (\mathsf{x}|\rho_{tot} (N \tau)| \mathsf{x} )\\
&= \sum_{\vec{m},\vec{m}'} \rho_{\vec{m},\vec{m}'} \sigma(\mathsf{x}-S_{\vec{m}},\mathsf{x}-S_{\vec{m}'})\:,
\label{phiN}
\end{split}
\end{equation}
where $\vec{m} = (m_1,m_2, \ldots m_N)$ and 
 \begin{equation}
\begin{split}
  \rho_{\vec{m},\vec{m}'}& = U^N \mathcal{P}_{m_N}((N-1)\tau)  \ldots \mathcal{P}_{m_2}(\tau) \mathcal{P}_{m_1} \ini\\
&\quad \times    \mathcal{P}_{m_1'} \mathcal{P}_{m_2'}(\tau)\ldots \mathcal{P}_{m_N'}((N-1)\tau) U^{\dagger N}\:.
\end{split}
\end{equation}
Here $\mathcal{P}_{m}(k \tau) = U^{k\dagger} \mathcal{P}_{m} U^k$ denotes a time-evolved projection operator. For a Gaussian pointer state as defined in Eq.~\eqref{sig} the position matrix element can be expressed as
\begin{equation}
\begin{split}
\sigma(\mathsf{x} - S_{\vec{m}}, \mathsf{x} - S_{\vec{m}'}) &= g_{\sigma^2_\mathsf{x}}\big (\mathsf{x}-(S_{\vec{m}} +S_{\vec{m}'})/2 \big) \\
& \quad \times e^{-( S_{\vec{m}} -S_{\vec{m}'})^2/(8 \sigma^2_\mathsf{x})}\:.
\label{Smm}
\end{split}
\end{equation}
The shift of the Gaussian is determined by sums of those $N$ eigenvalues that are labelled by $\vec{m}$ and $\vec{m}'$ and hence read
\begin{equation}
S_{\vec{m}} = \sum_{k=1}^N \mu_{m_k}
\label{Sm}
\end{equation} 
and accordingly for the primed sequence $\vec{m}'$. 

As in Eq.~\eqref{p1} the pdf $P_N(\mathsf{x})$ of finding the pointer at $\mathsf{x}$ after $N$ contacts with the system is given by the trace over the non-normalized density matrix~\eqref{phiN} and hence becomes
\begin{equation}
\begin{split}
P_N(\mathsf{x}) &= \sum_{\vec{m},\vec{m}'} D^{\vec{m},\vec{m}'}_\ini g_{\sigma^2_\mathsf{x}} \big (\mathsf{x} -(S_{\vec{m}}  +  S_{\vec{m}'})/2 \big )\\
& \quad \times  e^{-(S_{\vec{m}}-  S_{\vec{m}'})^2/(8 \sigma^2_{\mathsf{x}}) }
\label{pNx}
\end{split}
\end{equation}
with the coefficients $D^{\vec{m},\vec{m}'}_\ini$ reading
\begin{equation}
\begin{split}
D^{\vec{m},\vec{m}'}_\ini &= \Tr\, \rho_{\vec{m},\vec{m}'}\\
&= \delta_{m_N,m_N'}\Tr\, \mathcal{P}_{m'_1} \mathcal{P}_{m'_2}(\tau) \ldots \mathcal{P}_{m_{N-1}'}\big ((N\!-\!2)\tau\big )\\
&\quad \times \mathcal{P}_{m_N}\big ((N\!-\!1)\tau \big) \ldots \mathcal{P}_{m_2}(\tau) \mathcal{P}_{m_1} \ini\:.
\label{Dmm}
\end{split}
\end{equation}   
These coefficients constitute the elements of a non-negative definite tensor of rank $N^2$ guaranteeing the positivity of the pdf $P_N(\mathsf{x})$ in spite of some of them being complex quantities. In view of the result of $N$ measurements discussed in the next section, we emphasize that the various Gaussian contributions to $P_N(\mathsf{x})$ all have the same variance $\sigma^2_\mathsf{x} = \langle Q^2 \rangle /\kappa^2$ resulting from the variance of the initial pointer state and the measurement strength parameter $\kappa$. In particular, the width of these contributions is independent of the number of measurements. For a sufficiently narrow width those contributions to the sum on the right hand side of Eq.~\eqref{pNx} stemming from vectors $\vec{m}$ and $\vec{m}'$ that lead to different sums $S_{\vec{m}}$ and $S_{\vec{m}'}$ are exponentially suppressed. Hence, if the inequality 
\begin{equation}
8 \sigma^2_\mathsf{x} \ll \min_{\vec{m},\vec{m}'\atop S_{\vec{m}} \neq S_{\vec{m}'} }   \big ( S_{\vec{m}} - S_{\vec{m}'} \big )^2 = \min_{m,m', \atop m \neq m'} \big (\mu_m -\mu_{m'} \big )^2
\label{ie} 
\end{equation}
is satisfied, then the $N$ contact strategy yields the statistics of the sums of eigenvalues of an observable $M$ read out at equal intervals of length $\tau$.
In this case one obtains as pdf of the sums the following expression
\begin{equation}
P_N(x) \approx \sum_{\vec{m},\vec{m'} \atop S_{\vec{m}} = S_{\vec{m}'}} D^{\vec{m},\vec{m}'}_\ini g_{\sigma^2_\mathsf{x}}(x-S_{\vec{m}})\:.
\end{equation} 
In the special case in which the system Hamiltonian $H_S$ and the observable $M$ commute, the projection operators $\mathcal{P}_m$' are constants of motion, $\mathcal{P}_m(t) = \mathcal{P}_m$ and the coefficients $D^{\vec{m},\vec{m'}}_{\ini}$ simplify to read
\begin{equation}
D^{\vec{m},\vec{m}'}_\ini = \prod_{k=1}^{N-1} \delta_{m_k,m_{k+1}} \delta_{m_k,m_k'}\:,
\end{equation}
yielding for the $N$-contact probability
\begin{equation}
P_N(\mathsf{x}) = \sum_m p_m\, g_{\sigma^2_\mathsf{x}}(\mathsf{x}-N \mu_m)
\label{PNc}
\end{equation}
with $p_m$ defined in Eq.~\eqref{pm}. This multiple-contact pdf resembles the single measurement pdf~\eqref{sp1} with the difference that all eigenvalues are multiplied by the number of contacts. The multiple-contact pdf is thus accordingly spread.

\section{multiple measurements}\label{mm}
In order to perform $N$ measurements of the same observable $M$ at equally spaced times $n \tau$, $n = 0,1, \dots (N-1)$ one may use the same number of equally prepared pointers, which are initially uncorrelated with each other as well as with the system. They are subsequently  brought in contact with the system and after the contact read out by a projective measurement. Consequently, the  non-normalized density matrix of the  system conditioned on the sequence of measurements $\vec{\mathsf{x}} \equiv (\mathsf{x}_1,\mathsf{x}_2, \ldots \mathsf{x}_N)$ takes the form
\begin{equation}
\begin{split}
{}^m\!\phi_{\vec{\mathsf{x}}}(\ini) &= \phi^1_{\mathsf{x}_N}(U \phi^1_{\mathsf{x}_{N-1}}(\ldots U \phi^1_{\mathsf{x}_1}(\ini) U^\dagger) \ldots U^\dagger)\\
&=\sum_{\vec{m},\vec{m}'} \rho_{\vec{m},\vec{m}'} \prod_{k=1}^N g_{\sigma^2_{\mathsf{x}}}(\mathsf{x}_k-(\mu_{m_k}+\mu_{m_k'})/2)\\
& \quad \times e^{-(\mu_{m_k} -\mu_{m_k'})^2/(8 \sigma^2_\mathsf{x})}\:.
\label{Mphix}
\end{split}
\end{equation}
The pdf ${}^m\!P_N(\vec{\mathsf{x}})$ to find the sequence $\mathsf{x}$ of measurement results becomes
\begin{equation}
\begin{split}
 {}^m\!P_N(\vec{\mathsf{x}})&= \Tr\, {}^m\!\phi^N_{\vec{\mathsf{x}}} \\
&= \sum_{\vec{m},\vec{m}'} D^{\vec{m},\vec{m}'}_\ini  \prod_{k=1}^N g_{\sigma^2_{\mathsf{x}}}(\mathsf{x}_k-(\mu_{m_k}+\mu_{m_k'})/2)\\
& \quad \times e^{-(\mu_{m_k} -\mu_{m_k'})^2/(8 \sigma^2_\mathsf{x})}\:.
\label{MpNxx}
\end{split}
\end{equation}
Hence, the pdf ${}^m\!P_N(\mathsf{x})$ to find the value $\mathsf{x}$ for the sum of the individual measurement results becomes
\begin{equation}
\begin{split}
{}^m\!P_N(\mathsf{x}) &= \int d^N \mathsf{x} \:\delta \big (\mathsf{x} - \sum_{k=1}^N \mathsf{x}_k \big ) {}^m\!P_N(\vec{\mathsf{x}})\\
&= \sum_{\vec{m},\vec{m}'} D^{\vec{m},\vec{m}'}_\ini g_{N \sigma^2_\mathsf{x}}(\mathsf{x}-(S_{\vec{m}}+S_{\vec{m}'})/2)\\
&\quad \times \prod_{k=1}^N e^{-(\mu_{m_k} - \mu_{m_k'})^2/(8 \sigma^2_\mathsf{x})}\:.
\label{MpNx}
\end{split}
\end{equation} 
As for the $N$-contact pdf~\eqref{pNx} the $N$-measurement pdf of the sum is a linear combination of Gaussians with centers at the same positions $(S_{\vec{m}} + S_{\vec{m}'})/2$ but with the $N$-fold variances. Hence, the $N$-contact pdf will in general have a much more detailed structure than the $N$-measurement pdf. 
\begin{figure}[t]
\includegraphics[width=\columnwidth]{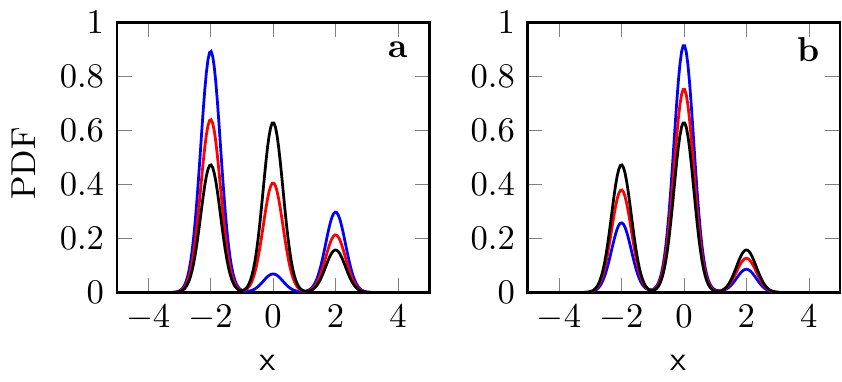}
\centering
\caption{\label{fig:0.5} (Color online) The pdfs $P_2(\mathsf{x})$ (blue solid lines) for unitary dynamics and ${}_dP_2(\mathsf{x})$ for dissipative dynamics at a dissipation rate (red and black solid lines) given by Eqs.~\eqref{qbp2} and~\eqref{dp2} characterizing unitary and dissipative dynamics. For unitary dynamics the relative weights of the central peak at $\mathsf{x} =0$ and the two side peaks $\mathsf{x} =\pm 2$ depend on the time between two contacts $\theta=B\tau = 11.7\pi/4$ in panel a and $\theta= 10.7\pi/4$ in panel b. 
In the presence of dissipation ($\gamma =(\Gamma_1+\Gamma_2)/B = 0.1$) the height ratio of the side peaks stays constant while they exchange weight with the  central peak (red lines). In the limit $r^d= \theta \gamma \to \infty$ the weights of the central and side peaks become equal (black curves an panels a and b).       
For all cases the initial density matrix is $\boldsymbol\tau_x =0.5$ yielding $p_1=\Tr \mathcal{P}_1\ini = 0.75$; further $q=\Tr \boldsymbol\tau_y\ini/2 = 0.43$, and $w = \Tr\, \boldsymbol\tau_z\ini /2=0$. The initial Gaussian state has variance $\sigma^2_{\mathsf{x}}=0.1$.}
\end{figure}

For a sufficiently small variance $\sigma^2_\mathsf{x}$ the last product-term in Eq.~\eqref{MpNx} suppresses all terms with $m_k \neq m_k'$. Hence only the diagonal part of the tensor $D^{\vec{m},\vec{m}'}_\ini$ contributes. For any observable having a non-degenerate spectrum it can be further simplified to read 
\begin{equation}
D^{\vec{m},\vec{m}}_\ini = \prod_{k=1}^{N-1} T(m_{k+1}|m_k) p_{m_1}
\label{Dmmsimp}
\end{equation}
where 
\begin{equation}
T(m|n) = |\langle m|U|n \rangle|^2
\label{Tmn}
\end{equation}
denotes the transition probabilities between eigenstates $|n \rangle$ and $|m \rangle$ of the observable $M$ governed by the unitary dynamics $U$. These probabilities form a bistochastic transition matrix of a Markovian chain~\cite{Cox} with the number of measurements specifying the chain length. The coefficients $D^{\vec{m},\vec{m}}_\ini$ are determined by the probability with which the sequence $\vec{m}$ starting at $m_1$ with probability $p_1$ occurs for this Markovian chain where the number of states that can be taken at each step equals the dimension $d_H$ of the Hilbert space of the system. 
For a large number of measurements such a Markovian chain typically  approaches a stationary regime in which the probability $1/d_H$ is assigned to all eigenvalues of the observable $M$. Hence the memory is lost of where the chain has started. Therefore those vectors $\vec{m}$ with a uniform distribution of elements $m_k$  acquire the highest probability for large $N$. 
One may expect that the average and the variance of the sums of $N$ eigenvalues asymptotically grow both in proportion to $N$ as in a normal random walk~\cite{Feller}. In exceptional cases the Markovian chain may cause strictly periodic trajectories which consequently also result in an asymptotically  periodic variation of the variance of the eigenvalue sum.       
\begin{figure}[t]
\includegraphics[width=\columnwidth]{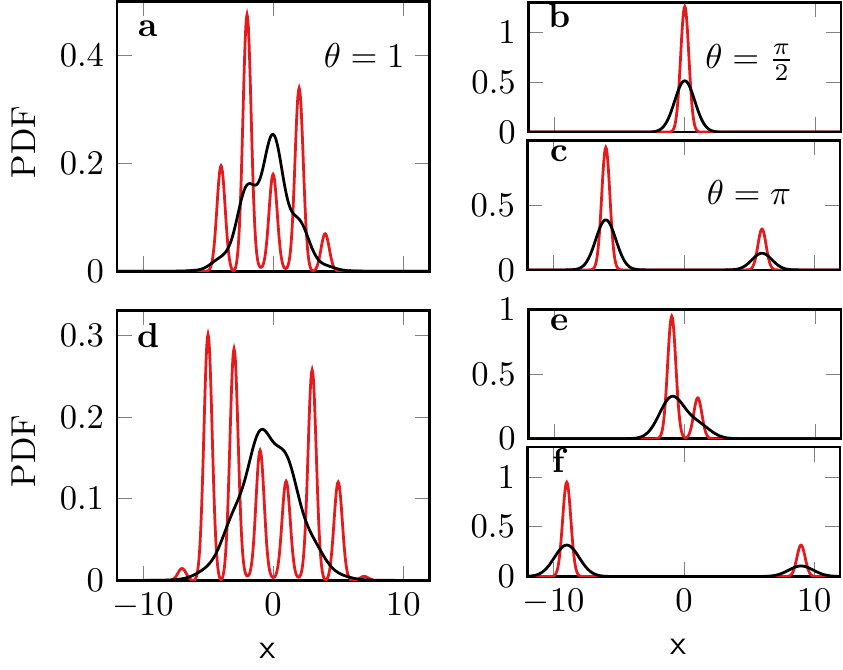}
\centering
\caption{\label{fig:1}(Color online) The pdf $P_N(\mathsf{x})$ given by Eq.~\eqref{pNx} of the pointer state after $N=6$ (panels a-c) and $N=9$ (panels d-f) contacts with a unitarily evolving qubit is displayed as a function of $\mathsf{x}$ (red solid lines) together with the according pdfs ${}^m\!P_N(\mathsf{x})$ [Eq.~\eqref{MpNx}] characterizing multiple measurements (black solid lines). The time between two contacts, as well as between two measurements, is $\theta = 1$ in panels a and d, $\theta = \pi/2$ in panels b and e, and $\theta= \pi$ in c and f. For all cases, the initial density matrix is determined by $\langle \boldsymbol\tau_x \rangle = 0.5$,  $\langle \boldsymbol\tau_y \rangle = 0$,  and $\langle \boldsymbol\tau_z \rangle = 0.78$. The initial Gaussian pointer state has variance $\sigma^2_\mathsf{x} =0.1$ leading for the contact scenario at the generic value $\theta =1$ to well separated lines centered at all but the two extreme positions of possible sums of the two eigenvalues ($\mu =\pm1$). The extreme positions have a too small weight to be visible.  For the exceptional value $\theta =\pi/2$ and $N=6$ there is only a single line centered at $\mathsf{x}=0$ and two lines for $N=9$ at  the positions of the eigenvalues, $\mathsf{x}=\pm 1$. For the other exceptional period $\theta =\pi$ two lines are located at $\mathsf{x} =\pm N$. The width of the individual lines is always determined by $\sigma_{\mathsf{x}}$ for repeated contacts. For repeated measurements the individual contributions merge to yield broad distributions. Only for $\theta = \pi$ the lines remain visible as they  are separated by $2N$ and the width is proportional to $\sqrt{N}$.}
\end{figure}

Another exceptional case occurs when the transition matrix agrees with the identity such as  
for observables commuting with the system Hamiltonian. Then one obtains with $T(m|n) = \delta_{m,n}$ the expression
\begin{equation}
{}^m\!P_N(\mathsf{x}) = \sum_m  p_m \,g_{N\sigma^2_\mathsf{x}}(\mathsf{x} - N \mu_m)\:.
\label{MpNc}
\end{equation}
leading, as for the  corresponding $N$-contact protocol, to a mixture of Gaussians at the positions of the observables eigenvalues, however with substantially enlarged variances. In both cases the variance of the eigenvalue sum grows as $N^2$.

\section{Dissipative dynamics}\label{dd}
In the previous sections the dynamics of the system between two consecutive contacts was considered to be unitary $U$. In most systems of practical interest ensuring perfectly unitary dynamics  is highly non-trivial and can only be achieved during a limited time span. In general, the implementation of the influence of an environment on the dynamics of a system in general poses a difficult problem. Here we assume weak coupling between the considered system and its environment resulting in a Markovian dynamics of the system described by a Lindblad master equation~\cite{Lindblad}. This dynamics maps the density matrix from an instant after a contact with the pointer on a new state time $\tau$ later by means of a linear, completely positive and trace preserving propagator $\mathcal{G}$. Hence, the initially factorizing density matrix of the total system after subsequent $N$ contacts and Markovian propagation of the system becomes
\begin{equation}
{}_d\rho_{tot}(N \tau) = (\mathcal{G V})^N(\ini\otimes \sigma)\:,
\label{drtot}
\end{equation}   
where $\mathcal{V}$ describes the action of a contact on the total density matrix which is given by
\begin{equation}
\mathcal{V}(\rho_{tot}) = V \rho_{tot} V^\dagger
\label{VVV}
\end{equation}
with the unitary contact operator $V$ defined in Eq.~\eqref{eq:3}. As the result of a projective measurement of the pointer position after $N$ contacts, one obtains as the non-normalized density matrix conditioned on the measurement result $\mathsf{x}$ an expression of the same structure as for a unitary dynamics, Eq.~\eqref{phiN}, reading for an initially Gaussian pointer state 
\begin{equation}
\begin{split}
{}_d\phi^N_x(\ini) &= \sum_{\vec{m},\vec{m}'} {}_d\rho_{\vec{m},\vec{m}'}\: g_{\sigma^2_\mathsf{x}}\!\big (\mathsf{x} -(S_{\vec{m}}  + S_{\vec{m}'} )/2 \big )\\ &\quad \times e^{-(S_{\vec{m}}  - S_{\vec{m}'} )^2/(8 \sigma^2_\mathsf{x})}\:,
\label{dphi}
\end{split}
\end{equation}
where the system operators ${}_d\rho_{\vec{m},\vec{m}'}$ in the presence of a Markovian dissipative dynamics are defined as
\begin{equation}
  \begin{split}
  {}_d\rho^{\vec{m},\vec{m}'} &= \mathcal{G}(\mathcal{P}_{m_N} \mathcal{G} ( \mathcal{P}_{m_{N-1}} \mathcal{G} ( \ldots \mathcal{P}_{m_2} \mathcal{G} (\mathcal{P}_{m_1} \ini \mathcal{P}_{m'_1})\\
  & \qquad \times \mathcal{P}_{m'_2} \ldots ) \mathcal{P}_{m'_{N-1}} ) \mathcal{P}_{m_N'} )\:.
  \label{D}
  \end{split}
\end{equation}
The probability ${}_dP_N(\mathsf{x})$ to find the result $\mathsf{x}$ again results from the trace of the non-normalized density matrix and hence becomes
\begin{equation}
\begin{split}
  {}_dP_N(\mathsf{x}) &= \sum_{\vec{m}, \vec{m}'} {}_dD^{\vec{m},\vec{m}'}_\ini  g_{\sigma^2_\mathsf{x}}\big (\mathsf{x} - (S_{\vec{m}}+S_{\vec{m}'})/2 \big )\\
&\quad \times e^{-(S_{\vec{m}}-S_{\vec{m}'})^2/(8 \sigma^2_\mathsf{x}) }\:.
\label{dpN}  
\end{split}
\end{equation}
It only differs from the above unitary result~\eqref{pNx} through the form of the coefficient matrix which is given by
\begin{equation}
\begin{split}
{}_dD^{\vec{m},\vec{m}'}_\ini & = \Tr\, {}_d\rho_{\vec{m},\vec{m}'}\\
&= \delta_{m_N,m_N'} \Tr\, \mathcal{P}_{m_N} \mathcal{G}(\mathcal{P}_{m_{N-1}} \mathcal{G}( \ldots  \mathcal{P}_{m_2}\\
&\quad \times \mathcal{G} (\mathcal{P}_{m_1} \ini \mathcal{P}_{m'_1})
 \mathcal{P}_{m'_2} \ldots ) \mathcal{P}_{m'_{N-1}} )\\ 
&= \delta_{m_N,m_N'} \Tr\, \mathcal{P}_{m_1'} \mathcal{G}^*(\mathcal{P}_{m_2'} \ldots \mathcal{G}^*(\mathcal{P}_{m_N})\\
&\quad  \ldots\mathcal{P}_{m_2}) \mathcal{P}_{m_1} \ini  
\label{dD}
\end{split}
\end{equation}
with $\mathcal{G}^*$ denoting the dual propagator satisfying $\Tr\, u \mathcal{G} (\chi) = \Tr \,\mathcal{G}^*(u) \chi$ for all bounded operators $u$ and all trace class operators $\chi$.  Assuming the validity of the quantum regression hypothesis~\cite{T86}, the last line can be interpreted as a multi-time correlation function $\langle \mathcal{P}_{m_1'} \mathcal{P}_{m_2'}(\tau) \ldots \mathcal{P}_{m_N}((N-1)\tau) \ldots  \mathcal{P}_{m_2}(\tau) \mathcal{P}_{m_1} \rangle$, in analogy to the expression~\eqref{Dmm} for unitary dynamics.
The Gaussians which are weighted by the above coefficients are located at the same positions and all have the same width as for an unitary dynamics.

Assuming that the Markovian dynamics asymptotically approaches a uniquely defined  stationary state $\rho^{st}$, the propagator acts as $\mathcal{G}(\chi) = \rho^{st} \Tr\, \chi$ on all trace class operators $\chi$, provided that the time $\tau$ between subsequent measurements is large enough. Under this condition the coefficients ${}_dD^{\vec{m},\vec{m}'}_\ini$ simplify considerably to read
\begin{equation}
 {}_dD^{\vec{m},\vec{m}'}_\ini=  \delta_{\vec{m},\vec{m}'}    
p_{\vec{m}}\:,
\label{Dst}
\end{equation}
where  $p_{\vec{m}} = p_{m_1} \prod_{k=2}^N p^{st}_{m_k}$ with $p_m =\Tr\, \mathcal{P}_m \ini$ and $p^{st}_{m_k} = \Tr\, \mathcal{P}_{m_{k}} \rho^{st}$ denotes the probability of finding the sequence of $N$  eigenvalues $\mu_{\vec{m}}$  whose first member is drawn from the initial distribution and all others are independently taken from the stationary distribution. Further we use as a shorthand  $\delta_{\vec{m},\vec{m}'}\equiv \prod_{k=1}^N \delta_{m_k,m'_k}$. 
Hence, the pdf ${}_dP_N(\mathsf{x})$ simplifies to read
\begin{equation}
{}_dP_N^{st}(\mathsf{x}) = \sum_{\vec{m}} g_{\sigma^2_\mathsf{x}} \big (\mathsf{x} - S_{\vec{m}} \big ) p_{\vec{m}}
\label{dPxst}
\end{equation}
Using the characteristic function of a Gaussian random variable given by $\int dx\, e^{i u x} g_{\sigma^2}(x-S) = e^{iu S} e^{-u^2 \sigma^2/2}$ one obtains the following expression for the characteristic function $G(u) = \int d\mathsf{x}\, {}_dP_N(\mathsf{x})\,e^{iu \mathsf{x}}$:
\begin{equation}
G(u) = e^{- u^2 \sigma^2_{\mathsf{x}}} \Tr\, e^{i u M} \!\ini \: \big( \Tr\, e^{iu M} \rho^{st} \big )^{N-1}\:.
\label{Gu}
\end{equation}
This expression, which is a product of the $N$ characteristic functions of the observable $M$ in the initial and subsequent stationary states, and of the characteristic function of the pointer position in its initial state, reflects the independence of the respective individual contributions to the total outcome.  Accordingly, the mean value ${}_d\langle \mathsf{x} \rangle = \int d\mathsf{x}\, {}_dP_N(\mathsf{x})\,\mathsf{x}$ and the variance ${}_d\Sigma^2_\mathsf{x} = \int d\mathsf{x}\, {}_dP_N(\mathsf{x})\,(\mathsf{x}-{}_d\langle \mathsf{x} \rangle)^2 $ result as
\begin{align}
{}_d\langle \mathsf{x} \rangle &= \langle M \rangle_0 + (N-1) \langle M \rangle_{st} \label{mvN}\\    
{}_d \Sigma^2_\mathsf{x} &= \sigma^2_\mathsf{x} + \langle \big ( M - \langle M \rangle_0 \big )^2 \rangle_0 \nonumber\\
& \quad + (N-1) \langle \big ( M - \langle M \rangle_{st} \big )^2 \rangle_{st} \label{varN}\:,
\end{align}
with $\langle M \rangle_0 = \Tr M\ini $ and $\langle M \rangle_{st} = \Tr M\rho^{st}$. For large values of contact numbers  the first moment as well as all cumulants  grow proportionally to $N$. Hence $\mathsf{x}$ behaves as a function of $N$ as a random walk. In particular, the contribution $\mathsf{x}/N$ per step acquires asymptotically a Gaussian distribution. This will remain true also as an asymptotic result for large $N$ if the time $\tau$ between two measurements is not large enough to lead to a complete approach to the stationary state. Due to its repeated action any dissipative dynamics leading to an uniquely defined stationary state will generate a Gaussian random walk-like behaviour after sufficiently many contacts.  

Finally we note that a protocol with $N$ measurements, as discussed in Sec.~\ref{mm}, in the presence of dissipation leads for the sum of measurements to an analogous expression as given in Eq.~\eqref{MpNx} with coefficients $D^{\vec{m},\vec{m}'}_\ini$ replaced by ${}_dD^{\vec{m},\vec{m}'}_\ini$. The main difference to the $N$-contact result~\eqref{dpN} is the broadening of the individual Gaussian contributions.

\begin{figure}[t] 
\includegraphics[width=\columnwidth]{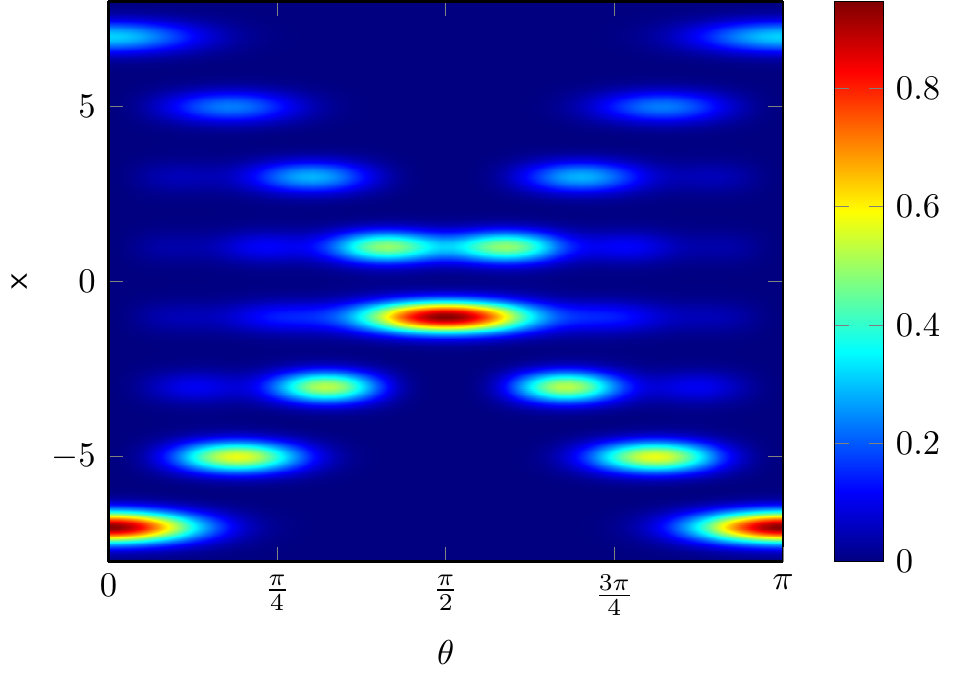}
\centering
\caption{\label{fig:2}(Color online) A color-coded presentation of the  pdf $P_7(\mathsf{x})$ as a function of the duration of the time $\theta$ and the pointer position $\mathsf{x}$. For generic values of $\theta$ the pdf exhibits local maxima at the possible values of the sums of seven eigenvalues $\pm 1$ at $-7,-5,\cdots, 5,7$. For particular times $\theta = 0, \pi$ the bimodal pdf~\eqref{eq:22} and for $\pi/2$ a unimodal pdf with maximum at $\mathsf{x}=0$ results. All distributions disclose a bias towards negative value due to the chosen initial density matrix which was chosen as specified in Fig.~\ref{fig:1}.}
\end{figure}
\section{Comparing the impact of contacts and measurements}
In order to quantify the average impact of repeated contacts on the state of the system we consider the reduced density matrix $\rho_k$ of the system immediately after the $k$-th contact and compare its trace distance with the density matrix $\rho(k \tau)$ which has evolved in the same time $k \tau$ in the absence of contacts. This distance is compared to the one of the density matrix after $k$ repeated measurements from the uninterrupted one $\rho(k \tau)$. The density matrix of the system after $k$ contacts is obtained by performing the partial trace over the pointer state of the total density matrix $\rho_{tot}(k \tau) = V (\mathcal{G V})^{k-1}(\ini\otimes \sigma) V^\dagger$, where $\mathcal{V}(\rho) =V \rho V^\dagger$. As above, $\mathcal{G}$ denotes the propagator of the Markovian dynamics between two contacts. In the case of unitary dynamics it acts as $\mathcal{G}(\rho) = U \rho U^\dagger $ see also Eq.~\eqref{UVN}. For a pointer initially staying in a Gaussian state, the partial pointer state trace can be performed to yield for the reduced density matrix 
\begin{equation}
\rho_k = \sum_{{}^k\!\vec{m},{}^k\!\vec{m}'} R^{{}^k\!\vec{m},{}^k\!\vec{m}'} e^{-\left [\sum_{j=1}^k (\mu_{m_j} -\mu_{m_j}') \right ]^2/(8\sigma^2_\mathsf{x}) }\:,
\label{rk}
\end{equation}
\begin{figure}[t]
\includegraphics[width=\columnwidth]{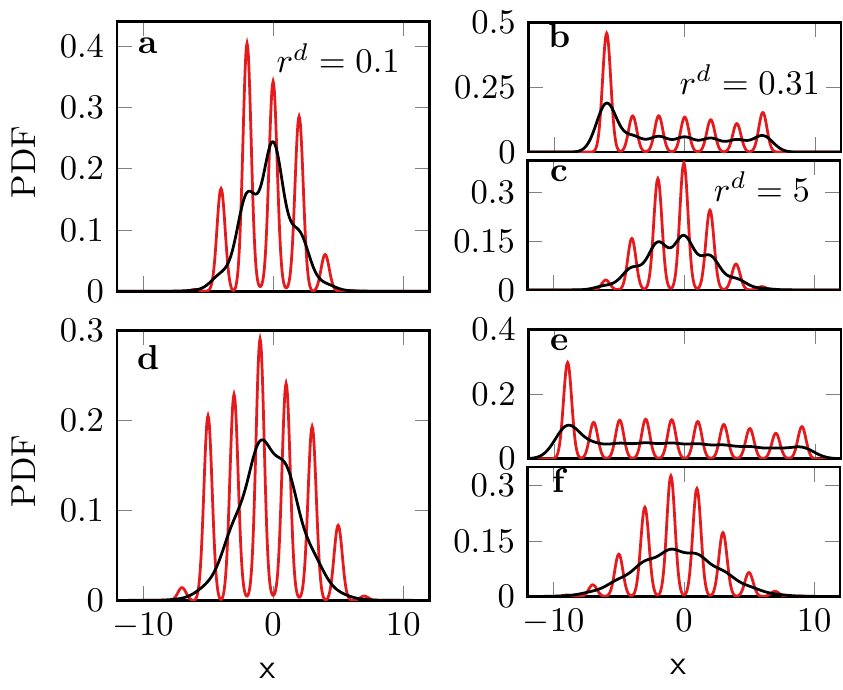}
\centering
\caption{\label{fig:4}(Color online) The multiple contact pdf  ${}_dP_N(\mathsf{x})$ (Eq.~\eqref{dpN}, red solid lines) and the repeated measurement pdf ${}_d^m P_N(\mathsf{x})$ (black solid lines) are displayed as functions of the pointer position $\mathsf{x}$ for various dissipation ratios $r^d = \gamma \theta = 0.1$ (panels a and d), $r^d= 0.31$ (panels b and e) and $r^d=5$ (panels c and f). Panels a-c are for $N=6$ and panels d-f are for $N=9$. The individual rates are chosen as $\Gamma_1/B =7.5\times 10^{-2}$ and $\Gamma_2/B= 2.5 \times 10^{-2}$. The initial density matrix of the qubit and the initial variance of the pointer are chosen as in the  Fig.~\ref{fig:1}.}
\end{figure}
where ${}^k\vec{m}$ denotes a vector whose number of components is $k$ and whose components are taken from the set indexing the eigenvalues of the tested observable $M$. Further, $R^{{}^k\!\vec{m},{}^k\!\vec{m}'}$ is a trace-class operator of the system indexed by a double series of left and right hand shifts of the pointer state. It is defined as
\begin{equation}
R^{{}^k\!\vec{m},{}^k\!\vec{m}'} 
=  \mathcal{P}_{m_k} \mathcal{G}(\mathcal{P}_{m_{k-1}} \ldots \mathcal{G}(\mathcal{P}_{m_1} \ini \mathcal{P}_{m_1'})
\mathcal{P}_{m_{k-1}'}) \mathcal{P}_{m_k'}\:.
\label{Rk}
\end{equation}
In contrast,  after $k$ non-selective measurements the reduced density matrix is given by
\begin{equation}
\rho^{m}_k  = \sum_{{}^k\vec{m},{}^k\vec{m}'} 
R^{{}^k\!\vec{m},{}^k\!\vec{m}'}
e^{ - \sum_{j=1}^k (\mu_{m_j} -\mu_{m_j}')^2/(8\sigma^2_\mathsf{x})}
\label{drmk}
\end{equation}
Both the $k$-contact and the $k$-measurement density matrices are linear combinations of the contact-specific operators $ R^{{}^k\!\vec{m},{}^k\!\vec{m}'}$, however being weighted by different coefficients. In the limit of a very wide initial pointer state, $\sigma^2_\mathsf{x} \to \infty$ these coefficients approach unity. With the completeness of the projection operators, $\sum_m \mathcal{P}_m = \mathbb{1}$ all sums can be performed and in both cases the back-action-free density matrix $(\mathcal{G})^{k-1} (\ini)$ results. In the other limit of precise measurements when the variance $\sigma^2_\mathsf{x}$ is larger than zero but satisfies the inequality~\eqref{ie}, the exponential factors on the right hand side of Eq.~\eqref{drmk} suppress all non-diagonal contributions after multiple measurements. In contrast, the non-diagonal elements and hence, coherences with respect to the eigen-basis of the observable $M$, are much less suppressed by the exponentiated  squared differences of the eigenvalue sums $S_{\vec{m}}$ compared to the exponentiated sums of squared eigenvalue differences.        

Finally, we note that in the limit of large $\tau$ for dissipative dynamics, such that stationarity is reached, one obtains the same result for multiple contacts and multiple measurements reading
\begin{equation}
\rho^{(m)}_k = \sum_{m_k,m_k'} \mathcal{P}_{m_k} \rho^{st}  \mathcal{P}_{m_k}  e^{-(\mu_{m_k} -\mu_{m_k'})^2/8 \sigma^2_{\mathsf{x}}}
\label{rmk}
\end{equation}
Note that this density matrix is independent of the number of contacts or measurements because the dynamics between contacts or measurements  erases any memory on the prehistory.

\section{Example: Qubit}\label{eqb}
In order to illustrate the  general theory outlined in the previous section we consider a quantum qubit as the system of interest whose Hamiltonian reads,
\begin{eqnarray}
\label{eq:20}
H &= \hbar B \boldsymbol\tau_z
\end{eqnarray}
with $\boldsymbol\tau_z$ being the $z$-component of the Pauli spin-$1/2$ matrix and $B$ being the strength of the Hamiltonian. We measure $\boldsymbol\tau_x$ for the qubit such that the operator $M = \boldsymbol\tau_x$. We first consider the case of unitary dynamics

\subsection{Unitary dynamics}
In spite of the fact that both the time-evolution of the free qubit, given by $U = \cos \theta -i \boldsymbol\tau_z \sin \theta$, and the spectral representation of $M$ with $\mu_1 =-1$, $\mathcal{P}_1 =(1 - \boldsymbol\tau_x)/2$ and  $\mu_2 =1$, $\mathcal{P}_2 =(1 + \boldsymbol\tau_x)/2$ are very simple, the exponential growth of the number of terms contributing to the pdf~\eqref{pNx} characterizing the $N$-contact protocol renders its analytic presentation for more than $N=2$ contacts basically impossible. Here, $\theta = B \tau$ specifies the duration of the unitary time-evolution between two contacts in units of the inverse frequency of the qubit. For $N=2$ one obtains after some lengthy algebra the expression    
\begin{equation}
\begin{split}
P_2(\mathsf{x}) &= \frac{1}{\sqrt{2\pi\sigma^2_\mathsf{x}}}\left [\left ( p_1 
e^{-\frac{(\mathsf{x}+2)^2}{2 \sigma^2_\mathsf{x}}} + p_2 e^{-\frac{(\mathsf{x}-2)^2}{2 \sigma^2_\mathsf{x}}} \right )\cos^2 \theta \right .\\ 
& \quad + e^{-\frac{\mathsf{x}^2}{2 \sigma^2_\mathsf{x}}} \sin^2 \theta \\
& \quad \left. - q \,e^{-\frac{1}{2\sigma^2_\mathsf{x}}}\left (e^{-\frac{(\mathsf{x}+1)^2}{2\sigma^2_\mathsf{x}}}- e^{-\frac{(\mathsf{x}-1)^2}{2\sigma^2_\mathsf{x}}}\right) \sin 2\theta  \right]
\label{qbp2}
\end{split}
\end{equation}
where $p_i = \Tr\, \mathcal{P}_i\ini $, as defined in Eq.~\eqref{pm}, $q = \Tr\, \boldsymbol\tau_y\ini /2$, and $w = \Tr\, \boldsymbol\tau_z\ini /2$ is restricted by $-\sqrt{p_1 p_2} \leq \sqrt{w^2+q^2} \leq \sqrt{p_1 p_2}$ because of the positivity of the initial density matrix $\ini$. For sufficiently small variances, say $\sigma^2_\mathsf{x} \lesssim 0.1$ one observes well separated peaks at the positions $\mathsf{x} = 0, \pm 2$.  At larger variances the peaks merge into a broad distribution. The contributions at $\mathsf{x} = \pm 1$ are never visible as peaks. At small variances their contribution is exponentially suppressed; at larger ones  they influence the form of the pdf as seen in Fig.~\ref{fig:0.5}. The  peak heights are governed by the probabilities $p_1$ and $p_2$ with which the eigenstates of the measured operator $\boldsymbol \tau_x$ with corresponding eigenvalues $\mu_1 =-1$ and $\mu_2 =1$, respectively, contribute to the initial density matrix $\ini$. Since we choose the state with $\mu=-1$ to have a higher occupancy the distribution is skewed to negative $\mathsf{x}$, a property that is carried forward to large $N$ as seen in Fig.~\ref{fig:1}.

Another analytic result emerges for the qubit system is when $\theta =n \pi$. In this case the time evolution operator $U$ commutes with the observable $M$ and hence yields with Eq.~\eqref{PNc} 
a bimodal pdf (Fig.~\ref{fig:1}c and f, red solid lines)
\begin{equation}
\label{eq:22}
P_N(\mathsf{x})|_{\theta = \pi} = \frac{1}{\sqrt{2 \pi\sigma^2_\mathsf{x}}} \left (  p_1 e^{-\frac{(\mathsf{x} + N )^2}{2\sigma^2_\mathsf{x}}}  + p_2e^{-\frac{(\mathsf{x} - N )^2}{2\sigma^2_\mathsf{x}}}  \right ).
\end{equation}
For this limiting case the average becomes $\langle \mathsf{x}\rangle = N(p_1-p_2)$ and the variance $\Sigma^2_{\mathsf{x}}=\langle \mathsf{x}^2\rangle -\langle \mathsf{x}\rangle^2 = \sigma^2_\mathsf{x} + 2 p_1 N^2$. The repeated measurement case [Eq.~\eqref{MpNx}, Fig.~\ref{fig:1}c and f, black solid lines] can also be analytically evaluated for $\theta = \pi$  resulting in a similar expression as above with $\sigma^2_{\mathsf{x}} \rightarrow N^2\sigma^2_{\mathsf{x}}$.

In general, the pdf for a repeated contacts $P_N(\mathsf{x})$ displays a complex behaviour as shown by red solid lines in Fig.~\ref{fig:1}a and d. As expected, performing repeated measurements significantly broadens the pdf and we loose all individual measurement related information even for a small number of measurements [see Fig.~\ref{fig:1}a and d]. 

Yet another special case turns out to be $\theta= \pi/2$. Because the time evolution flips the projection operators $\mathcal{P}_m(n \pi/2) = {P}_{(-1)^nm}$ for $n$ odd, and leaves them unchanged  for $n$ even, a unimodal pdf  being centered at $\mathsf{x}=0$ emerges for any even number of contacts ($N=6$, see Fig.~\ref{fig:1}b) while it becomes bimodal with a higher weight at negative values of $\mathsf{x}$ (due to the initial $\ini$ with $p_1> p_2$) for an odd number of contacts ($N=9$, see Fig.~\ref{fig:1}e). The behaviour as a function of $\theta$ is captured in Fig.~\ref{fig:2} displaying a perfect bimodal distribution occurring at $\theta = 0,\pi/2,\pi$. At all other values intermediate peaks of the distribution are visible but much weaker as compared to the dominant peak at negative $\mathsf{x}$ skewing the average to negative values of $\mathsf{x}$.

\begin{figure}[t] 
\includegraphics[width=\columnwidth]{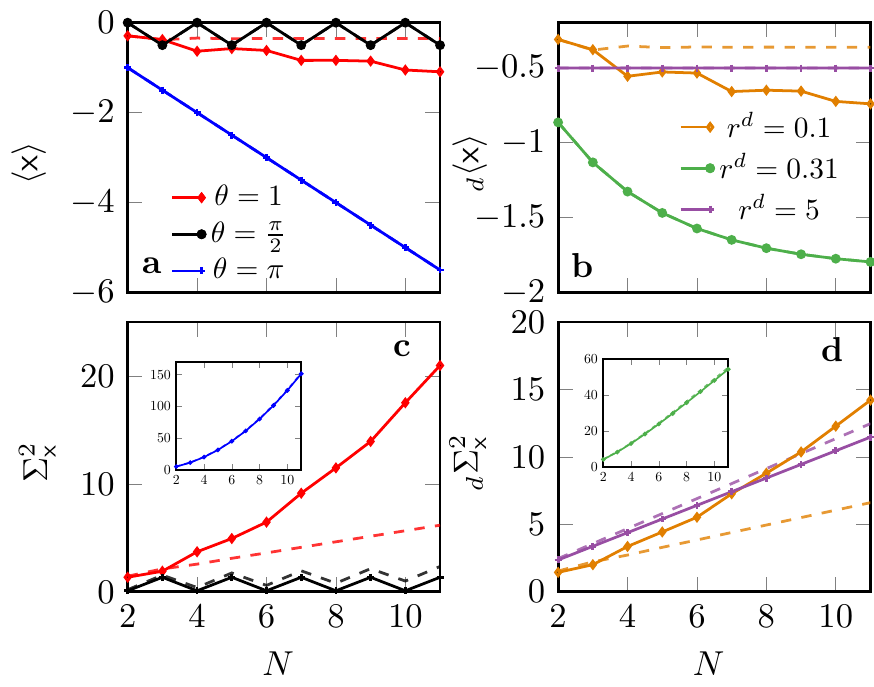}
\centering
\caption{\label{fig:3}(Color online) The mean values $\langle\mathsf{x}\rangle$ (panels a and b) and the variances $\Sigma^2_{\mathsf{x}}$ (panels c and d) of the pointer position as a function of the number of repeated contacts (solid lines) and the number of repeated measurements (dashed lines) for unitary (panels a and c) and for dissipative dynamics (panels b and d). 
For unitary dynamics with the atypical duration $\theta=\pi/2$ the oscillation of the pdf between a bimodal and a unimodal form leads to an oscillatory behaviour of the mean and an $N$ independent variance (black lines in panels a and c) for repeated contacts. 
For repeated measurements the mean follows the same oscillatory behavior with the variance of each Gaussian peak scaling with $N$ giving $\Sigma^2_{\mathsf{x}}$ a weak dependence on $N$. For $\theta = \pi$ both the repeated contact and measurement schemes yield with Eqs. (\ref{PNc}) and (\ref{MpNc}), respectively, a linear increase of the absolute mean value $\langle \mathsf{x} \rangle$ and a quadratic increase  of the variance $\Sigma^2_\mathsf{x}$ as displayed in the inset of panel c. For typical values of $\theta$, the repeated contact scheme with unitary dynamics gives rise to a qualitatively similar, but less pronounced ballistic diffusion behavior. In contrast, repeated measurements lead to a saturation of the mean value and a linear growth of the variance, i.e. to normal diffusion of $\mathsf{x}$. A transition to normal diffusion is also observed in the presence of dissipation. For the largest dissipation with $r^d = 5$ the mean value and the variance are in good agreement with the Eqs.~(\ref{mvN}) and (\ref{varN}). Other parameters are chosen as in Fig.~\ref{fig:1}.}
\end{figure}

We restrict the discussion of repeated measurements for cases with a sufficiently narrow initial pointer state variance $\sigma^2_\mathsf{x}$. The  pdf ${}^m\!P_N(\mathsf{x}) $ characterizing $N$ measurement then contains only diagonal coefficients $D^{\vec{m},\vec{m}}_\ini$ which are determined by a transition matrix with the elements $T(1|1) =T(2|2) = \cos^2 \theta$ and $T(1|2) = T(2|1) =\sin^2 \theta$, see Eqs.~(\ref{Dmmsimp}) and (\ref{Tmn}). With an increasing number of $N$ the relatively narrow  lying lines typically merge resulting in a broad pdf with a most probable value at $\mathsf{x} \approx 0$. Accordingly, the variance grows proportional to $N$ for all generic values of $\theta$, such as $\theta=1$, see the black curves displayed in Fig.~\ref{fig:1} panels a and d. The exceptional cases   $\theta=n\pi$, $n=0,1,2$  yield the identity for the transition matrix. According to Eq.~(\ref{MpNc}) the pdf ${}^m\!P_N(\mathsf{x})$ displays two separate lines at $\pm N$ each of which has a width proportional to $N$ (see Figs.~\ref{fig:1}c,d and \ref{fig:3}a,c). Consequently, the variance $\Sigma^2_\mathsf{x}$ increases as $N^2$. On the other hand the choice $\theta=\pi/2, 3\pi/2$ leads to a periodic Markov chain which entails an alternatingly uni- and bimodal pdf for even and odd $N$, respectively. Then also the variance is periodic (see also Fig.~\ref{fig:3}  a,c). 

\subsection{Dissipative dynamics}
Next we discuss the case of dissipative dynamics which is assumed to be governed by a Lindblad-type master equation reading
\begin{equation}
\begin{split}
\dot{\rho}(t) &= -i B [\boldsymbol\tau_z, \rho(t)] + \Gamma_1 \big ( [\boldsymbol\tau_-,\rho \boldsymbol\tau_+] +[\boldsymbol\tau_-\rho, \boldsymbol\tau_+] \big )\\
&\quad +\Gamma_2 \big ( [\boldsymbol\tau_+,\rho \boldsymbol\tau_-] +[\boldsymbol\tau_+\rho, \boldsymbol\tau_-] \big )\:,
\label{me}
\end{split}
\end{equation}
where $\Gamma_1 > \Gamma_2$. The operators $\boldsymbol\tau_z$,  $\boldsymbol\tau_+ = (\boldsymbol\tau_x + i \boldsymbol\tau_y)/2$, $\boldsymbol\tau_- = (\boldsymbol\tau_x - i \boldsymbol\tau_y)/2$ and $\mathbb{1}$ form a complete set which transforms under the dual propagator as
\begin{equation}
\begin{split}
\mathcal{G}^*(\boldsymbol\tau_+) &= e^{(2i - \gamma) \theta} \boldsymbol\tau_+\:,\\
\mathcal{G}^*(\boldsymbol\tau_-) &= e^{-(2i + \gamma) \theta} \boldsymbol\tau_- \:,\\
 \mathcal{G}^*(\boldsymbol\tau_z) &= e^{-2 \gamma \theta} \boldsymbol\tau_z + \frac{\Gamma_1 - \Gamma_2}{\Gamma_1 + \Gamma_2} \mathbb{1} \:,\\
 \mathcal{G}^*(\mathbb{1}) &= \mathbb{1} \:.
\label{ttt}
\end{split}
\end{equation}
Here, the dimensionless damping rate is defined as $\gamma =(\Gamma_1 + \Gamma_2)/B$. 
Similarly as in the unitary case we can find the analytic form of the distribution when two measurements are made ($N=2$)
\begin{equation}
\begin{split}
{}_dP_2(\mathsf{x}) &= \frac{1}{\sqrt{2 \pi \sigma^2_\mathsf{x}}} \left [ \left( p_1 e^{-\frac{(\mathsf{x}+2)^2}{2 \sigma^2_\mathsf{x}}}   +p_2 e^{-\frac{(\mathsf{x}-2)^2}{2 \sigma^2_\mathsf{x}}}  \right ) f_c(\theta) \right.\\
& \quad + e^{-\frac{\mathsf{x}^2}{2 \sigma^2_\mathsf{x}}} f_s(\theta)\\
& \quad \left . -q  e^{-\frac{1}{2 \sigma^2_\mathsf{x}}} \left ( e^{-\frac{(\mathsf{x}+1)^2}{2 \sigma^2_\mathsf{x}}} -e^{-\frac{(\mathsf{x}-1)^2}{2 \sigma^2_\mathsf{x}}} \right ) e^{-\gamma \theta} \sin 2 \theta \right ]
\label{dp2}
\end{split}
\end{equation}
with 
\begin{equation}
\begin{split}
f_c(\theta) &= (1+e^{-\gamma \theta}\cos 2\theta)/2,  \\
f_s(\tau) &= (1-e^{-\gamma \theta}\cos 2\theta)/2.
\label{fcs}
\end{split}
\end{equation}
In absence of dissipation, Eq.~\eqref{dp2} maps exactly to Eq.~\eqref{qbp2}. The presence of any finite dissipation exponentially reduces the influence of the Gaussian contributions at $\mathsf{x} = \pm 1$. Further it changes the relative weights of the peaks at $\mathsf{x}=\pm 2$ and $\mathsf{x}=0$: The side peaks increase if $\cos 2 \theta >0$ (Fig.~\ref{fig:0.5}a blue solid line) while the central peak grows if $\cos 2 \theta <0$ (Fig.~\ref{fig:0.5}b blue solid line). In the limit of large times $\theta \to \infty$ the asymptotic result following from Eq.~\eqref{dPxst} for $N=2$ is approached (Fig.~\ref{fig:0.5} black solid lines). This is because the stationary density matrix resulting from the master equation~\eqref{me} is diagonal with respect to the $\boldsymbol\tau_z$ eigen-basis such that both stationary probabilities $p^{st}_{1,2}$ have the same value $1/2$ independent of the stationary expectation value $\langle \boldsymbol\tau_z \rangle_{st} = (\Gamma_1 -\Gamma_2)/ (\Gamma_1 +\Gamma_2)$.   

For two larger values of $N$ some results are presented in Fig.~\ref{fig:4}. For the small dissipation parameter $r^d\equiv \gamma \theta = 0.1$ the system has not enough time to thermalize between measurements and hence the pdfs displayed in the Fig.~\ref{fig:1}a and d resemble those for unitary dynamics. At the intermediate value of $r^d= 0.31$ additional two peaks at the maximal positions $\mathsf{x}=\pm N$ become visible with maximal weight at $-N$ while all other peaks are of approximately the same height, see Fig.~\ref{fig:1}b and e. With  a further increase of the dissipation parameter the system between two measurements is driven into the  stationary state yielding the asymptotic result Eq.~\eqref{dPxst} for the pdf. It takes a further simplified form for any initial density matrix that commutes with $\boldsymbol\tau_z$ such that, with $p_{\vec{m}} =(1/2)^N$,   the pdf assumes the form of a mixture of Gaussians with binomial weights. The centers of these Gaussians are located at the possible values taken by the sums of all combination of eigenvalues. The  binomial weights reflect the number of different combinations of $N$ eigenvalues with the same sum. In this particular case the pdf hence reads
\begin{equation}
{}_dP^{st}_N(\mathsf{x}) = \left(\frac{1}{2} \right )^N\sum_{k=0}^N\binom{N}{k} g_{\sigma^2_\mathsf{x}}(\mathsf{x} - N + 2 k)\:.
\label{dpxst}
\end{equation} 
In spite of the fact that the initial state has an off-diagonal matrix element with respect to the $\boldsymbol\tau_z$-basis, the agreement of the results displayed in Fig.~\ref{fig:4}c and f is very good.        

A rough characterization of the $N$-dependence of the statistics of $\mathsf{x}$, specifying either the number of  contacts or of measurements, both for the unitary and the dissipative case is provided by the mean-value and the variance, $\langle \mathsf{x} \rangle$ and $\Sigma^2_\mathsf{x}$ as displayed for a few cases in Fig.~\ref{fig:3}. 
Typically, the mean value grows proportionally to the number of contacts as well as to the number of measurements both for unitary and for dissipative dynamics as illustrated by the upper two panels of Fig.~\ref{fig:3}. An exception from this rule occurs for $\theta =\pi/2$  displaying oscillations of the mean value for unitary dynamics in agreement with the above described behavior of the underlying pdfs alternating as a function of $N$ between the same unimodal and bimodal shapes.   

The influence of dissipation suppresses correlations between the shifts of the pointer states at contacts that are separated by a sufficiently large dissipation ratio $r^d$ leading to a variance asymptotically growing proportionally to the number of contacts displaying the characteristic feature of normal diffusion. For the here considered most simple qubit, and as we expect also for other so-called integrable quantum systems~\cite{Haake}, the variance increases with $N^2$ as in the case of classical, ballistic diffusion. This observation though is based on our numerical results which are restricted to a relatively low number of contacts. One might speculate that for unitary chaotic dynamics the variance still grows super-diffusively, but governed by a power law with an exponent between one and two. The different behaviour of unitary and dissipative dynamics for a qubit is illustrated in Fig.~\ref{fig:3}c and d, respectively.

\subsection{Trace distance}\label{td}
\begin{figure}[t]
\includegraphics[width=\columnwidth]{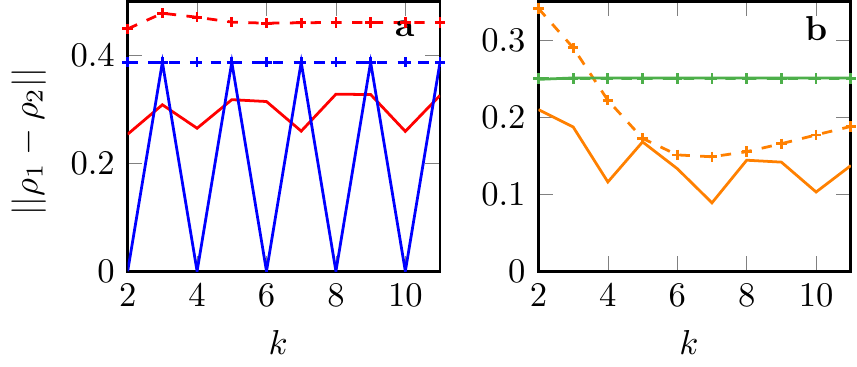}
\centering
\caption{\label{fig:5}(Color online) The trace distances $||\rho_{1} -\rho_2||$ according to Eq.~\eqref{eq:td} from a density matrix $\rho_2\equiv\rho(k \theta)$ uninterruptedly evolved up to the time $k \theta$, either by unitary (panel a) or dissipative dynamics (panel b), to the corresponding dynamical scheme $\rho_1$ with $k$ repeated contacts (solid lines) or $k$ measurements (dashed lines). In the case of unitary dynamics the time between contacts or measurements is $\theta=1$ (red lines) and $\theta= \pi/2$ (blue lines). In the case of dissipative dynamics results are displayed for the dissipation rates $r^d =0.1$ (orange lines) and $r^d =5$ (green lines). Typically, the distance between the interrupted and uninterrupted scenario is smaller for repeated contacts than for repeated measurements indicating a less severe back-action by the contacts compared to the measurements. Exceptions are found for unitary dynamics at $\theta=\pi/2$  with an odd number of interruptions and for the case  of the large dissipation rate leading to an almost perfect approach to the stationary state of the qubit. The remaining parameters are chosen as in Fig.~\ref{fig:1}. The lines connecting the points at integer values of $k$ are meant to guide the eye.}
\end{figure}
In Fig.~\ref{fig:5} the trace-distances between the density matrices of systems affected by a number of contacts or measurements and those freely propagating are compared as a function of the number of contacts. Here we use the trace distance between two density matrices $\rho_\alpha$ ($\alpha =1,2$), describing qubits having the expectation values $\langle \boldsymbol\tau_k \rangle^\alpha =\Tr\,\boldsymbol\tau_k\rho_\alpha$ ($k=x,y,z$), given by
\begin{equation}
|| \rho_1 - \rho_2|| = \Big [ \sum_k \big (\langle \boldsymbol\tau_k \rangle^1 - \langle \boldsymbol\tau_k \rangle^2 \big )^2 \Big ]^{1/2}\:.
\label{eq:td}
\end{equation}
It turns out that the trace-distance is typically smaller for the less invasive repeated contact scenario (solid lines) than for repeated measurements (dashed lines) in Fig.~\ref{fig:5}. In case of repeated contacts with unitary dynamics in between with $\theta =\pi/2$ (blue lines in Fig.~\ref{fig:5}a), the odd number of contacts and measurements result in the same trace distance because then one finds operators $R^{{}^k\!\vec{m}, {}^k\!\vec{m}'} = \prod_{l=1}^k \delta_{m_l,m_l'} \mathcal{P}_{m_k}\rho\mathcal{P}_{m_k} $ yielding equal exponential weights in the Eqs.~\eqref{rk} and~\eqref{drmk}. The other exception from the rule is for complete equilibration where the $k$ contacts and $k$ measurements lead to the same density matrix, see Eq.~\eqref{rmk} [blue lines in Fig.~\ref{fig:5}b].    

\section{Conclusions}\label{con}
We investigated in some detail a possibility to gain information about the values of an observable taken at subsequent times with as little back-action on the system as possible. The primary information on the system is taken within intervals of time that are negligibly short; it is transferred to the state of a pointer where their subsequent contributions are accumulated.         
After each contact the system is allowed to move freely, i.e. without being influenced by the pointer. The pointer itself is assumed to be idle until it is contacted again. The final readout of the pointer in terms of a projective measurement yields a value that coincides within some error margin with the sum of the observable at the instants of contacts. Other values, corresponding to the algebraic mean of two such sums are exponentially suppressed as long as the mentioned error margin is narrow enough. 

The resulting statistics of the final pointer state may be interpreted in terms of discrete quantum walks \cite{qw}. Here, the walker is realized by the pointer and the system performs as the coin deciding in which direction and how far the walker moves in a step. In the example of a qubit the Hilbert space of the coin has dimension two as frequently assumed in the theory of quantum walks but generalizations to coins living in arbitrarily large Hilbert spaces are straightforward. Depending on the kind of dynamics of the system the resulting random walk may vary from ballistic, i.e., with a variance growing proportionally to the square of the number of steps, to normal diffusion with a linear growth of the variance. The latter behavior is found for dissipative dynamics of the system governed by a Markovian master equation. For the qubit, undergoing unitary dynamics, the persistent correlations of the system dynamics apparently lead to a ballistic behavior for the relatively small numerically accessible numbers of contacts. The behavior of the variance for large numbers of contact as well as for more complex systems undergoing unitary dynamics poses an interesting problem and might provide a novel way to characterize so-called quantum chaotic systems. Presently, the investigation of this problem is hampered by numerical problems because it requires both large system Hilbert spaces as well as a large number of contacts. Both demands request huge storage and computational capacities which can possibly be realized with future quantum computers.

We would like to emphasize that the proposed strategy to reduce the back-action by repeated contacts differs from  weak measurements specifying the probabilities of   so-called quantum-trajectories of a continuously measured observable \cite{Wiseman, GWM, JS} in several respects: Instead of a continuum of necessarily weak measurements determining a quantum trajectory we consider a discrete sequence of contacts which are not restricted to be weak. The final measurement then yields a random number representative for the sum of the sequence of the observable taken at the contacts.      

We are also aware of the fact that the experimental realization of a continuous pointer which is idle if not in contact with the system might be difficult. We presented the approach of repeated contacts intentionally with an idealized model in order not to hide the principle idea with  technical complications being specific for a particular realistic application.   

As a particular problem that can be attacked by the presented strategy we finally mention the diagnosis of a quantum engine performing in finite time. The so far employed analysis in terms of projective energy measurements \cite{Ding} suppresses any coherences extending over several cycles. Their possible impact on the performance with respect to power, efficiency and reliability is of major importance \cite{Plenio17}.      
 
\section*{Acknowledgements}
This research was supported by the Institute for Basic Science in Korea (IBS-R024-Y2).


\begin{thebibliography}{100}
\bibitem{op} An operation is a linear, completetely positive map on the trace-class operators of a Hilbert space~\cite{Kraus, Hayashi}.
\end{thebibliography}
\end{document}